\documentclass[aps,prl,superscriptaddress,preprint,amsmath,amssymb]{revtex4}
\usepackage{graphicx}

\begin{document}

\title{Effect of an in-plane magnetic field on the interlayer phase coherence in the extreme-2D organic superconductor $\kappa$-(BEDT-TTF)$_2$Cu(NCS)$_2$}

\author{A. E. Kovalev}
\affiliation{Department of Physics, University of Florida,
Gainesville, FL 32611,USA}
\author{S. Takahashi}
\affiliation{Department of Physics, University of Florida,
Gainesville, FL 32611,USA}
\author{S. Hill}
\email[corresponding author, Email:]{hill@phys.ufl.edu}
\affiliation{Department of Physics, University of Florida,
Gainesville, FL 32611,USA}
\author{J. S. Qualls}
\affiliation{Department of Physics, Wake Forest University,
Winston-Salem, NC 27109, USA}

\date{\today}

\begin{abstract}
Using a high-sensitivity cavity perturbation technique (40 to
180~GHz), we have probed the angle dependent interlayer
magneto-electrodynamic response within the vortex state of the
extreme two-dimensional organic superconductor
$\kappa$-(BEDT-TTF)$_2$Cu(NCS)$_2$. A previously reported
Josephson plasma resonance [M. Mola {\em et al}., Phys. Rev. B
{\bf 62}, 5965 (2000)] exhibits a dramatic re-entrant behavior for
fields very close ($<1^\circ$) to alignment with the layers. In
this same narrow angle range, a new resonant mode develops which
appears to be associated with the non-equilibrium critical state.
Fits to the angle dependence of the Josephson plasma resonance
provide microscopic information concerning the superconducting
phase correlation function within the vortex state. We also show
that the effect of an in-plane magnetic field on the temperature
dependence of the interlayer phase coherence is quite different
from what we have previously observed for perpendicular magnetic
fields.

\smallskip


\bigskip

\noindent{Corresponding Author:}
\newline
Dr. Stephen Hill
\newline
Department of Physics
\newline
University of Florida
\newline
Gainesville, FL 32611 (USA)
\newline
FAX: 1-352-392-3591
\newline
Email: hill@phys.ufl.edu

\end{abstract}

\maketitle

\clearpage

\noindent{{\bf 1. Introduction}}

\noindent{The quasi-two-dimensional (Q2D) organic superconductor
$\kappa$-(BEDT-TTF)$_2$Cu(NCS)$_2$ (BEDT-TTF denotes
bis-ethylenedithio-tetrathiafulvalene, or ET for short) represents
a model system for studying vortex physics in layered systems and,
in particular, the influence of vortices on the interlayer phase
coherence within the mixed state~\cite{IshiguroYamaji}. Like the
high temperature superconductors (HTS),
$\kappa$-(ET)$_2$Cu(NCS)$_2$ possesses a highly anisotropic
layered structure with the superconducting ET planes separated by
insulating anion layers~\cite{IshiguroYamaji}. The anisotropy
parameter $\gamma$, defined as the ratio of the interlayer
(currents~$\parallel a-$axis) and in-plane (currents~$\parallel
bc-$plane) penetration depths ($\gamma \equiv
\lambda_a/\lambda_{bc}$), is thought to be in the range
$50-200$~\cite{CarringtonPRL99}, {\em i.e}. similar to that of
Bi$_2$Sr$_2$CaCu$_2$O$_{8+d}$~\cite{MatsudaPRL97}. In contrast to
the HTS, however, organic conductors are extremely clean, with
very few crystal defects. Also, because of the reduced T$_c$
($\sim 9-10$ K) and B$_{c2}$ ($\sim 5$~T for fields $\perp$
layers, and $30-35$~T for fields $\parallel$ layers), one can
probe much more of the temperature/field parameter space within
the superconducting state than is currently possible in the HTS.}

Our recent investigations of the microwave response of
$\kappa$-(ET)$_2$Cu(NCS)$_2$ in the 30 to 200~GHz range have shown
that a Josephson Plasma Resonance (JPR) dominates the interlayer
electrodynamics within the mixed
state~\cite{MolaPRB00,HillJPCM02}. The JPR frequency, $\omega_p$,
depends explicitly on the Josephson coupling (interlayer phase
coherence) between layers; indeed, $\omega_p$ is directly related
to the spatial and temporal average of the function $\langle \cos
\varphi_{n,n+1}(${\bf r}$) \rangle$, where $\varphi_{n,n+1}(${\bf
r}) is the gauge invariant difference in the phase of the order
parameter between layers $n$ and $n+1$, at position {\bf r} within
the layers. Thus, the JPR serves as an extremely sensitive tool
for probing the structure of vortex phases in this and other
layered organic and inorganic
superconductors~\cite{KoshelevPRB00}. Application of a magnetic
field ($B_z$) normal to the layers introduces pancake vortices
into the sample, accompanied by large intra-layer phase
fluctuations. Due to the weak nature of the Josephson coupling in
the title compound, the pancake vortices in adjacent layers become
decoupled at relatively weak applied fields $(\sim 10$~mT). This
decoupling transition dramatically suppresses the interlayer phase
coherence $\langle \cos \varphi_{n,n+1}(${\bf r}$) \rangle$ which,
in turn, suppresses $\omega_p$. Theoretical and experimental
studies within the vortex liquid phase have shown that the field
and temperature (T) dependence of $\omega_p$ may be described as

\begin{equation}
\omega _p^2 \left( {B_z ,T} \right) \propto B_z^{ - \nu } T^{ - 1}
\end{equation}
\

\noindent{with $\nu$ being slightly less than
unity~\cite{MolaPRB00,HillJPCM02,KoshelevPRB00}.}

Until recently, our investigations were limited to the
perpendicular field geometry. These studies revealed a
transformation from a pinned vortex phase, to a depinned liquid
state~\cite{MolaPRB00,HillJPCM02}. Our improved experimental setup
now allows angle dependent microwave impedance measurements of
small single crystal samples (see below). Here, we report
preliminary data mainly within the vortex liquid phase for fields
close to alignment with the $bc-$plane ($\parallel$~highly
conducting layers). Application of an in-plane field ($B_x$)
introduces an additional term in the phase difference
$\varphi_{n,n+1}(${\bf r}) due to the vector potential associated
with $B_x$. This leads to a further suppression of $\omega_p$,
which is directly related to the Fourier transform of the phase
correlation function $S(${\bf r}$) = \langle \cos
\varphi_{n,n+1}(${\bf r}$)-\cos \varphi_{n,n+1}($0$)
\rangle$~\cite{KoshelevPRB00}. Thus, angle dependent JPR
measurements provide a direct means of extracting microscopic
parameters associated with the superconducting state.

\bigskip

\noindent{{\bf 2. Experimental}}

\noindent{Measurements were performed using a sensitive cavity
perturbation technique described elsewhere~\cite{MolaRSI}. As a
source and detector, we use a millimeter-wave vector network
analyzer (MVNA), enabling phase sensitive measurements covering
the frequency range from 40 to $\sim200$~GHz. This phase
sensitivity allows us to determine both components of the complex
sample response~\cite{MolaRSI}, though we only plot the
dissipative response in this article. A single crystal sample was
placed within a cylindrical cavity such that a combination of
in-plane and interlayer currents were excited; a detailed
discussion of the electrodynamics is published
elsewhere~\cite{HillRC}. Interlayer currents penetrate deep
($100~\mu$m$-1$~mm) into the sample due to the low conductivity in
this direction. Dissipation then depends on the ratio of the
interlayer and in-plane conductivities, as well as the relative
dimensions of the sample. For this geometry, we have shown
unambiguously that dissipation is dominated by the interlayer
conductivity for the highly anisotropic organic
conductors~\cite{MolaRSI,HillRC}.}

A single $\kappa$-(ET)$_2$Cu(NCS)$_2$ crystal, with approximate
dimensions $0.7\times0.5\times0.2$~mm$^3$, was used for these
investigations. Temperature control in the 2 to 10~K range was
maintained using a variable flow cryostat (Quantum Design PPMS),
with Cernox resistance thermometers acting as temperature sensors.
Smooth rotation of the entire rigid microwave probe, relative to
the horizontal DC field produced by a 7~T split-pair magnet, was
achieved via a room temperature stepper motor mounted at the neck
of the magnet dewar; the stepper motor offers $0.1^\circ$ angle
resolution. The source and detector are bolted rigidly to the
microwave probe; subsequent connection to and from the MVNA is
achieved via flexible coaxial cables~\cite{MolaRSI}. In this mode
of operation, one can maintain optimal coupling between the
spectrometer and the cavity containing the sample, {\em whilst}
rotating the probe. As discussed in great detail in
ref.~\cite{MolaRSI}, good coupling between the various microwave
elements is essential in order to maintain a high sensitivity and
a low noise level. All data presented in this paper were obtained
whilst sweeping the applied magnetic field at different fixed
angles $\theta$ relative to the least conducting $a^*$ direction
($\perp bc-$plane).

\bigskip

\noindent{{\bf 3. Results and discussion}}

\noindent{Figure~1 plots microwave dissipation versus magnetic
field, at various angles $\theta$; the temperature was 4.3~K and
the frequency 71.36~GHz for these measurements. The main broad
peak seen at all angles corresponds to the JPR which we have
reported previously~\cite{MolaPRB00,HillJPCM02}, {\em i.e}. this
mode displays all of the previously published behavior at $\theta
= 0^\circ$. The JPR peak position ($B_{res}$) follows an
approximately $1/\cos\theta$ dependence up to
$\theta\approx80^\circ$, whereupon it reaches a maximum, tending
towards zero-field as $\theta\rightarrow90^\circ$ (see dashed
curve). This re-entrant behavior of the JPR may be attributed to
the in-plane field component ($B_x$), as discussed in the
introduction. In this same narrow angle range ($\theta >
80^\circ$), a new low-field resonance (LFR) develops (dotted
curve), which moves to higher fields as $\theta \rightarrow
90^\circ$. The LFR exhibits a dramatic hysteretic behavior wherein
it is only observed for up-sweeps of the magnetic field. This is
illustrated in Fig.~2, where we plot the temperature dependence
(raw data) of the JPR and LFR for both up- and down-sweeps of the
magnetic field, with $\theta$ close to $90^\circ$ ($\pm
0.1^\circ$). The data in Fig.~2 data were obtained during a
separate run to the data in Fig.~1, {\em i.e}. the sample is the
same, but was subjected to thermal cycling to room temperature
between experiments. Furthermore, the frequency of 74.20~GHz is
slightly different. Nevertheless, angle dependent studies revealed
an identical behavior to the data in Fig.~1 (see Fig.~3).}

We first discuss the re-entrant behavior of the JPR at angles
close to $\theta = 90^\circ$. An analytic expression for the angle
dependence within the vortex liquid phase may be obtained by
assuming a Gaussian form for $S(${\bf r}$)$, giving

\begin{equation}
\omega _p^2  \approx \omega _p^2 \left( {B_z ,T} \right)\exp
\left( { - \frac{{\pi s^2 B_x^2 }}{{\Phi _o B_z }}} \right)
\propto \frac{1}{{B\cos \theta }}\exp \left( { - \frac{{\pi s^2
B\sin ^2 \theta }}{{\Phi _o \cos \theta }}} \right)
\end{equation}
\

\noindent{where {\em s} is the interlayer spacing, and $\Phi_o$ is
the flux quantum~\cite{KoshelevPRB00}. The dashed curve in Fig.~3
represents a fit to the above expression for the data obtained
from the first experiment (open squares). While such a fit is
reasonable, a better fit (solid curve in Fig.~3) involves a slight
modification to the denominator of the pre-exponential factor in
Eq.~(2), such that $B \cos\theta \rightarrow B [\cos\theta +
\alpha]$. The fit assumes fixed $\omega_p = 2\pi f$ and $B =
B_{res}$, and yields $\alpha = 0.1$ and a reasonable value for $s
\sim 8$~$\AA$. The modification to Eq.~(2), which produces a
rounding of the $1/\cos\theta$ behavior close to $\theta = 90^o$,
may have several explanations: i) it may reflect a limitation in
the Gaussian approximation for $S(${\bf
r}$)$~\cite{KoshelevPRB00}; or ii) it may indicate that the
pancake vortex contribution to $\omega_p$ is not determined solely
by $B_z$. The latter may suggest that a highly anisotropic 3D
picture (as opposed to strictly 2D) offers the more realistic
description of the superconductivity in the title compound, or
this could also be indicative of paramagnetic pair breaking for
large in-plane fields. Both of these effects have been considered
in the context of other angle dependent studies of the
superconductivity in $\kappa$-(ET)$_2$Cu(NCS)$_2$, such as
$H_{c2}$\cite{ZuoPRB00}. The obtained value for {\em s} is
somewhat smaller than the actual interlayer spacing. However, the
exponent in Eq.~(1) is approximate, {\em i.e}. it assumes that the
phase correlation length is approximately equal to the intra-layer
inter-vortex separation, through $B_z$.}

Next we turn to the hysteretic behavior, and the temperature
dependence of the JPR and LFR seen in Fig.~2. The hysteresis is
clearly a manifestation of the irreversible region of the vortex
phase diagram, which extends beyond the maximum field sweep range
at the lowest temperature of 3~K; in fact, the JPR is barely seen
for the up-sweep at this temperature. Eq.~(2) was derived for a
vortex liquid. Therefore, the re-entrance of the JPR at angles
close to $90^\circ$ drives it into the irreversible region of the
phase diagram at low temperatures (see Fig.~4 and
ref.~\cite{LangPRB94}); hence, its disappearance. Nevertheless,
for all other temperatures and angles, the data presented here for
the JPR were confined to the reversible liquid-like
phase~\cite{HillJPCM02}. On the other hand, the LFR resides
entirely within the irreversibility line~\cite{LangPRB94}. We note
that, although the JPR is not very pronounced at angles close to
$90^\circ$ (Fig.~2), we do always see two distinct resonances on
the up-sweeps (this is very clear in Fig.~1).

We now speculate that the LFR also corresponds to a JPR which is
associated with the non-equilibrium critical state; the origin of
its opposing angle dependence is not presently known. Indeed, the
tendency to move to higher fields implies increasing interlayer
phase coherence as $\theta\rightarrow90^\circ$ (see discussion in
ref.~\cite{HillJPCM02}), which is quite contrary to the
expectations of Eq.~(2). Looking at the temperature dependence of
the resonance positions for angles close to $\theta = 90^\circ$
(Fig.~4), one finds the same opposing trends for the two
resonances, {\em i.e}. the JPR moves to higher fields with
increasing temperature, indicating increased interlayer phase
coherence, while the LFR exhibits the opposite behavior. Thus,
within the irreversible region of the phase diagram, and for large
in-plane fields ($\theta$ close to $90^\circ$), the interlayer
phase coherence decreases with increasing temperature, {\em i.e}.
$\partial\langle \cos \varphi_{n,n+1}(${\bf r}$)
\rangle/\partial$T~$< 0$; the opposite holds for the reversible
phase $(\partial\langle \cos \varphi_{n,n+1}(${\bf r}$)
\rangle/\partial$T~$> 0$). These trends are entirely contrary to
what we have previously observed with the applied field
perpendicular to the layers, $(\theta =
0^\circ)$~\cite{HillJPCM02}. This could be related to the
anisotropy of the superconductivity, {\em i.e}. the
phenomenological constant $\alpha$ in our fit. However, further
experimental and theoretical studies are clearly needed in order
to understand this behavior, particularly in the irreversible
phase.

\bigskip

\noindent{{\bf 4. Summary and conclusions}}

\noindent{Within the vortex liquid phase, angle dependent
electrodynamic investigations of $\kappa$-(ET)$_2$Cu(NCS)$_2$
reveal a re-entrance of the JPR frequency, $\omega_p$, for fields
close to alignment with the highly conducting layers. We are able
to fit the angle dependence to a single analytic expression which
is related to the superconducting phase correlation function. A
sharp, low field, resonance is also observed in the vicinity of
$\theta = 90^\circ$, which appears to be associated with the
non-equilibrium critical state.}

\bigskip

\noindent{{\bf 5. Acknowledgements}}

\noindent{This work was supported by NSF (DMR0196430, DMR0239481
and DMR0196461). SH is a Cottrell scholar of the Research
Corporation.}




\begin{thebibliography}{10}

\bibitem{IshiguroYamaji}
T. Ishiguro, K. Yamaji, and G. Saito, {\em Organic
Superconductors}, Vol.~88 of
  {\em Springer Series in Solid State Sciences} (Springer-Verlag, Berlin,
  1998).

\bibitem{CarringtonPRL99}
A. Carrington, I.~J. Bonalde, R. Prozorov, R.~W. Giannetta, A.~M.
Kini, J.
  Schlueter, H.~H. Wang, U. Geiser, and J.~M. Williams, Phys. Rev. Lett. {\bf
  83},  4172  (1999), and references therein.

\bibitem{MatsudaPRL97}
Y. Matsuda, M.~B. Gaifullin, K. Kumagai, M. Kosugi, and K. Hirata,
Phys. Rev.
  Lett. {\bf 78},  1972  (1997).

\bibitem{MolaPRB00}
M. Mola, J.~T. King, C.~P. McRaven, S. Hill, J.~S. Qualls, and
J.~S. Brooks,
  Phys. Rev. B {\bf 62},  5965  (2000).

\bibitem{HillJPCM02}
S. Hill, M.~M. Mola, and J.~S. Qualls, J. Phys. Condens. Matter
{\bf 14},  6701
   (2002).

\bibitem{KoshelevPRB00}
A.~E. Koshelev, L.~N. Bulaevskii, and M.~P. Maley, Phys. Rev. B
{\bf 62},
  14403  (2000).

\bibitem{MolaRSI}
M. Mola, S. Hill, P. Goy, and M. Gross, Rev. Sci. Inst. {\bf 71},
186  (2000).

\bibitem{HillRC}
S. Hill, Phys. Rev. B {\bf 62},  8699  (2000).

\bibitem{ZuoPRB00}
F. Zuo, J.~S. Brooks, R.~H. McKenzie, J.~A. Schlueter, and J.~M.
Williams,
  Phys. Rev. B {\bf 61},  750  (2000).

\bibitem{LangPRB94}
M. Lang, F. Steglich, N. Toyota, and T. Sasaki, Phys. Rev. B {\bf
49},  15227
  (1994).

\end{thebibliography}



\clearpage

\noindent{{\bf Figure captions}}

\bigskip

\noindent{FIG. 1. Angle dependence of the microwave dissipation
due to interlayer currents; the field orientations are indicated
in the figure. Only up-sweeps are shown (see also Fig.~2). The
temperature is 4.3~K and the frequency is 71.36~GHz.}

\bigskip

\noindent{FIG. 2. Temperature dependence (raw data) of the JPR and
LFR for both up- and down-sweeps of the magnetic field, with
$\theta$ close to $90^\circ$ $(\pm 0.1^\circ)$. The frequency is
74.20~GHz, and the temperatures are indicated in the figure.}

\bigskip

\noindent{FIG. 3. A plot of the JPR and LFR positions (in field)
versus field orientation ($\theta$). JPR data are included for two
experimental runs, the first at a frequency of 71.36~GHz, the
second at a frequency of 74.20~GHz; the temperature was 4.3~K for
the first run, and 3.5~K for the second run. The dashed curve is a
fit to Eq.~(2) for the JPR data obtained in the first run; the
solid curve is a modified fit (see text).}

\bigskip

\noindent{FIG. 4. A plot of the temperature dependence of the JPR
and LFR resonance positions (in field), with $\theta$ close to
$90^\circ$ $(\pm 0.1^\circ)$. The frequency is 74.20~GHz. The
dashed curve represents the irreversibility line estimated for
this field orientation on the basis of published data (see open
squares, and ref.~\cite{LangPRB94}).}

\clearpage
\begin{figure}

\includegraphics{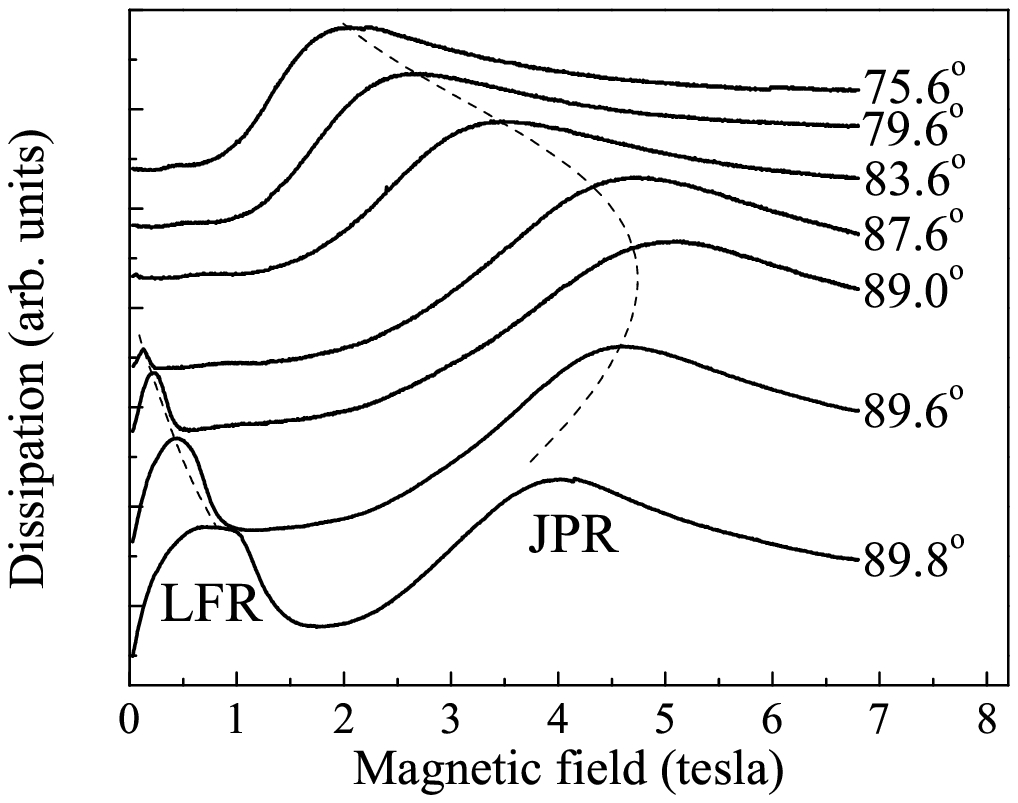}
\caption{\label{Fig1} Kovalev {\em et al.}}
\end{figure}

\bigskip

\begin{figure}

\includegraphics{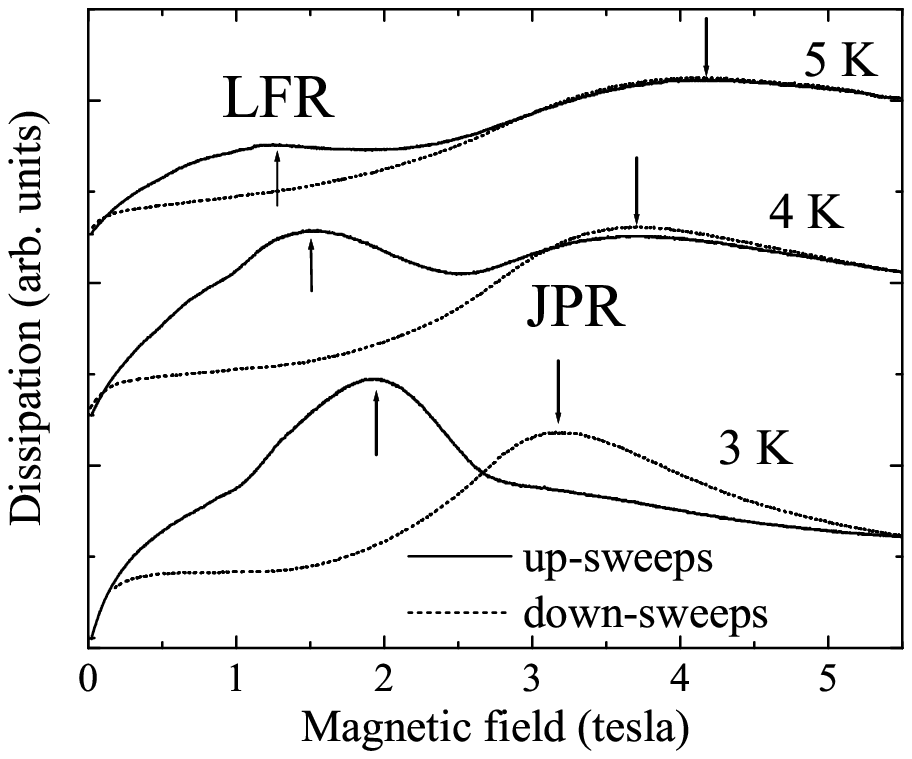}
\caption{\label{Fig2} Kovalev {\em et al.}}
\end{figure}

\bigskip

\begin{figure}

\includegraphics{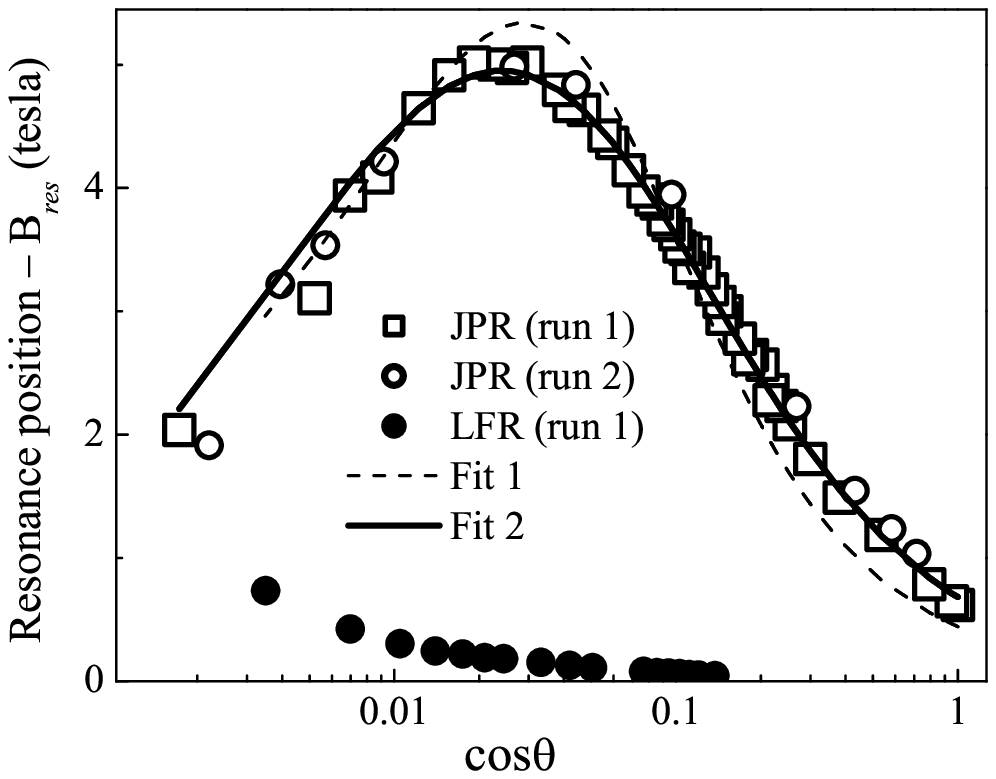}
\caption{\label{Fig3} Kovalev {\em et al.}}
\end{figure}

\bigskip

\begin{figure}

\includegraphics{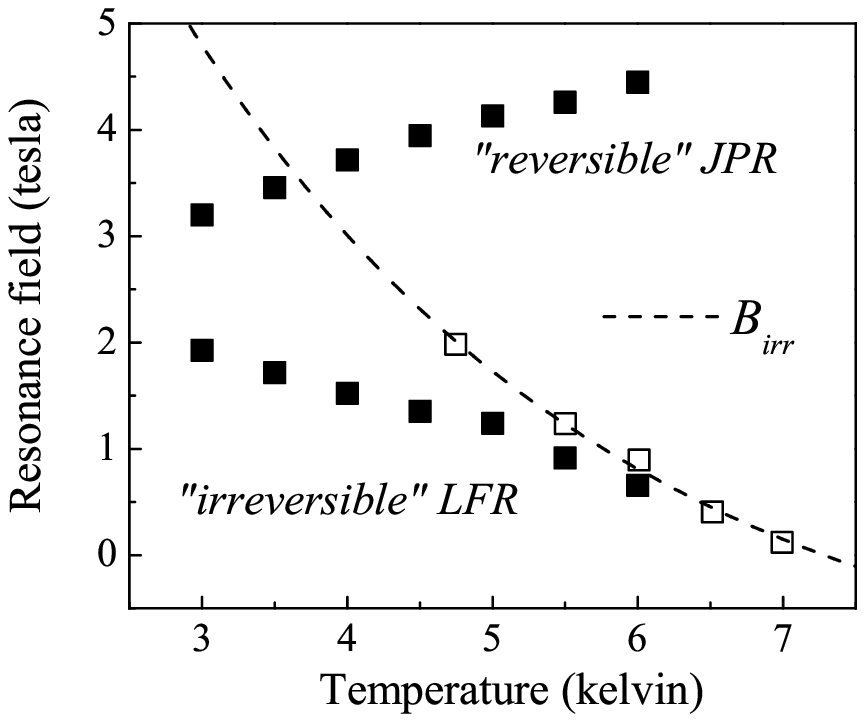}
\caption{\label{Fig4} Kovalev {\em et al.}}
\end{figure}

\end{document}